\documentclass[doublecol]{epl2}
\title{Accuracy in strategy imitations promotes the evolution of fairness in the spatial ultimatum game}
\shorttitle{Accuracy in strategy imitations promotes the evolution of fairness in the spatial ultimatum game}

\author{Attila Szolnoki,\inst{1} Matja{\v z} Perc,\inst{2} Gy{\"o}rgy Szab{\'o}\inst{1}}
\shortauthor{Szolnoki et al.}

\institute{\inst{1}Institute of Technical Physics and Materials Science, Research Centre for Natural Sciences, Hungarian Academy of Sciences, P.O. Box 49, H-1525 Budapest, Hungary\\
\inst{2}Faculty of Natural Sciences and Mathematics, University of Maribor, Koro{\v s}ka  cesta 160, SI-2000 Maribor, Slovenia}


\abstract{Spatial structure has a profound effect on the outcome of evolutionary games. In the ultimatum game, it leads to the dominance of much fairer players than those predicted to evolve in well-mixed settings. Here we show that spatiality leads to fair ultimatums only if the intervals from which the players are able to choose how much to offer and how little to accept are sufficiently fine-grained. Small sets of discrete strategies lead to the stable coexistence of the two most rational strategies in the set, while larger sets lead to the dominance of a single yet not necessarily the fairest strategy. The fairest outcome is obtained for the most accurate strategy imitation, that is in the limit of a continuous strategy set. Having a multitude of choices is thus crucial for the evolution of fairness, but not necessary for the evolution of empathy.}

\begin{document}

\maketitle

After having discovered the conflict between individual and common interests that emerges amongst selfish agents, which today is widely known as the prisoner's dilemma, Merrill Flood asked his colleagues how they would share the profit of the second-hand dealer when directly buying a car from each other \cite{poundstone_92}. It turned out that the majority seller-buyer pairs agreed on a fair split. This characteristic human behavior outlines how social dilemmas could be resolved.

The ultimatum game was proposed by G{\"u}th \textit{et al.} \cite{guth_jebo82} with the aim of studying the experimental outcome of ultimatum bargaining. In addition to the traditional goal of bargaining, which is to solve a distribution problem between the involved parties \cite{binmore_92}, ultimatum bargaining concerns the case where one party can restrict the set of possible agreements to a single proposal with which the other party can either agree or not. Besides the many applications in experimental economics, there is a rapidly growing interest in the ultimatum game stemming from many different fields of research, which is fueled by the game's fascinating ability to capture the most fundamental aspects of sharing that takes place in the absence of contracts and regulatory institutions \cite{sigmund_sa02}. The rules of the game can be laid down with just a couple of sentences. Imagine two players having to share a sum of money. One proposes a split, and the other can either agree with it or not. No haggling is allowed. If there is an agreement, the sum is shared according to the proposal. If not, both players remain
empty handed.

Similarly simple as the rules of the ultimatum game is the decision each rational proposer ought to make, which is to claim the large majority of the sum, given that the responder ought to
accept even the smallest amount offered as otherwise she would receive nothing. This scenario is indeed predicted by theoretical economics under the assumption that each individual is fully
rational and focused only on maximizing its own profit. Experiments, however, reveal a different reality. Largely regardless of sex, age and the amount of money at stake, people
refuse to accept offers they perceive as too small \cite{camerer_03, thaler_jep88, guth_jep90, roth_aer91, bolton_geb95, henrich_s10, henrich_pnas12}. Offers below one third of the total amount are rejected as often as they are accepted, and not surprisingly, more than two thirds of all offers will be remarkably close to the fair 50:50 split. It thus turns out that humans are remarkably fond of fair play, regardless of whether it is enforced with contracts and regulations or not \cite{sigmund_sa02}. The question is why?

There is psychological evidence suggesting that the utility functions based on which we determine our satisfaction are not simply the reflections of our own profits, but the profits of others as well \cite{kirchsteiger_jebo94, bethwaite_jep96}. While it is not completely clear how we weight them, the origins of such behavior might stem from the fact that we want to be members of strong groups, on which we can rely on for protection and help with rearing offspring \cite{nowak_11, hrdy_11}. This is of course impossible if we don't share our rewards equally with the other members of the group. An altogether different explanation is that we offer equal shares because we fail to reap the benefits of a one-off encounter. Put differently, we fail to ``seize the moment'' although it should be clear to us that the probability of another encounter with the same player is minute \cite{fehr_qje99}.

Similarly to previous observations concerning social dilemmas \cite{pacheco_jtb06, perc_pre08, fu_pre08b, pacheco_ploscb09, sigmund_dga11, du_f_dga11, pinheiro_pone12, gomez-gardenes_jtb08, vukov_pone11}, theoretical works have emphasized the importance of reputation \cite{nowak_s00}, empathy \cite{page_bmb02, sanchez_jtb05}, spatial structure \cite{page_prsb00, killingback_prsb01, iranzo_jtb11, deng_ll_pa11}, and heterogeneity \cite{da-silva_r_bjp07, da-silva_r_jtb09} for the successful evolution of fairness. In particular, Page \textit{et al.} \cite{page_prsb00} have shown that in well-mixed populations natural selection favors the rational solution, while spatiality may lead to much fairer outcomes. This result has been tested thoroughly against different types of players and updating rules \cite{iranzo_jtb11}, on various interaction networks \cite{kuperman_epjb08, eguiluz_acs09, li_x_pre09, sinatra_jstat09, xianyu_b_pa10b}, as well as under coevolution \cite{gao_j_epl11}.

\begin{figure}
\begin{center}
\includegraphics[width = 8cm]{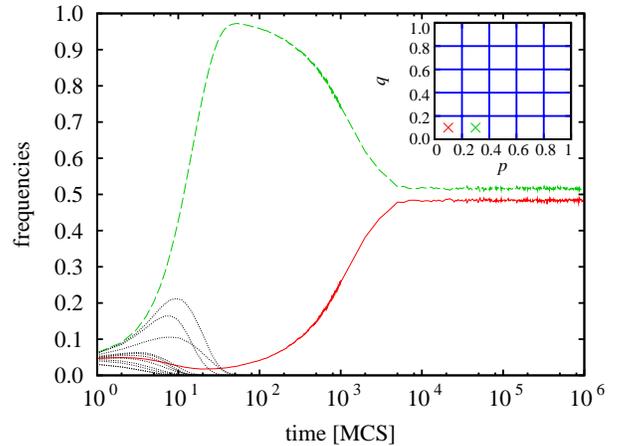}
\caption{\label{time}Time evolution of the fraction of strategies as obtained for $N=5$ and $L=400$. The surviving strategies, $i=j=0$ and $i=j+1=1$, are marked by red (solid) and green (dashed) lines. Inset shows the schematic presentation of the coarse-grained $p-q$ strategy plane for $N \times N = 5^2$ strategies. Crosses mark the two coexisting strategy cells that survive the evolutionary process in this particular case.}
\end{center}
\end{figure}

Here we depart from the traditionally studied version of the ultimatum game by no longer considering the unit intervals from where the players choose their offer level $p$ and acceptance level $q$ to be continuous, but rather, we consider both intervals to be coarse grained. The inset of Fig.~\ref{time} depicts such a setup schematically if the number of intervals along $p$ and $q$ is equal to $N=5$. In \cite{szolnoki_ug1}, we have introduced and studied a similar variant of the ultimatum game, yet with an imposed $p=q$ constraint and an invading strategy in order to reveal the fascinatingly rich dynamical behavior that underlies human bargaining. It encompasses mixed stationary states and traveling waves, as well the spontaneous emergence of cyclic dominance, which we have shown to be a consequence of pattern formation. In this letter, however, we focus on the survivability of strategies in the complete $p-q$ plane with the aim of determining the impact of a finite number of choices when possing the ultimatum.

Initially, we thus have $N^2$ strategies $E_{i,j}$, where $i,j=0, 1, \ldots, N-1$, that are distributed uniformly at random on a $L \times L$ square lattice. Each player on site $x$ is assigned a $(p_x,q_x)$ pair that corresponds to its strategy $E_{i,j}$ such that $p_x=(r_1+i)/N$ and $q_x=(r_2+j)/N$, where the two real random numbers $(r_1,r_2) \in [0,1)$ are drawn independently for the creation of each $E_{i,j}$. The evolution of the initial strategy distribution is performed by repeating the following elementary steps in accordance with the Monte Carlo simulation procedure. First, a randomly selected player $x$ acquires its payoff $U_x$ by playing the ultimatum game with its four nearest neighbors, whereby during each pairwise interaction acting once as a proposer with $p_x$ and once as a responder with $q_x$, according to its $E_{i,j}$ strategy. Next, a randomly chosen neighbor $y$ with strategy $E_{i^\prime,j^\prime}$ also acquires its payoff $U_y$ in the same way. Lastly, player $x$ tries to enforce its $E_{i,j}$ strategy on player $y$ in accordance with the probability $w=\{1+\exp[(U_{y}-U_{x})/K]\}^{-1}$, where $K$ quantifies the uncertainty during the strategy adoption process. We emphasize that the adoption of a new strategy is possible only if $i\neq i^\prime$ and/or $j \neq j^\prime$. Furthermore, during the imitation of a strategy $E_{i,j}$ ($i,j=0, \ldots, N-1$) two new independent random numbers ($r_1$ and $r_2$) are generated for the corresponding value of $p_y$ and $q_y$ in order to take into account the role of mutation, which was also considered in the continuous ultimatum game \cite{nowak_s00}.

There is another advantage to using the adoption of $(i,j)$ indices rather than copying the $(p_x,q_x)$ values accurately. In the latter case, it might happen that the final state will depend on the finite number of choices when posing the ultimatum, especially at small system size when only a limited set of $(p,q)$ values is available. To generate new $(p_x,q_x)$ values after strategy adoption, however, diminishes the fortuity of the initial state while preserving the course of strategy evolution. We also note that our analysis is restricted only to odd values of $N$ (from $N=3$ to $321$), which ensures that the fair $p=q=0.5$ pair is positioned in the center of the strategy $i=j=(N-1)/2$. In this way, artifacts related to the coexistence of neighboring strategies along the $p=q=0.5$ border can be avoided.

During the evolutionary process the time is measured in Monte Carlo steps (MCS). During one MCS each player has a chance once on average to change its strategy. For the systematic numerical analysis, we have determined the fraction of strategies in the final stationary state when varying the coarse graining $N$ at a fixed noise level ($K=0.1$). To ensure an adequate accuracy, we have used $L=200-1600$, and we have averaged the final state over $10^2-10^4$ independent runs.

\begin{figure}
\begin{center}
\includegraphics[width = 4cm]{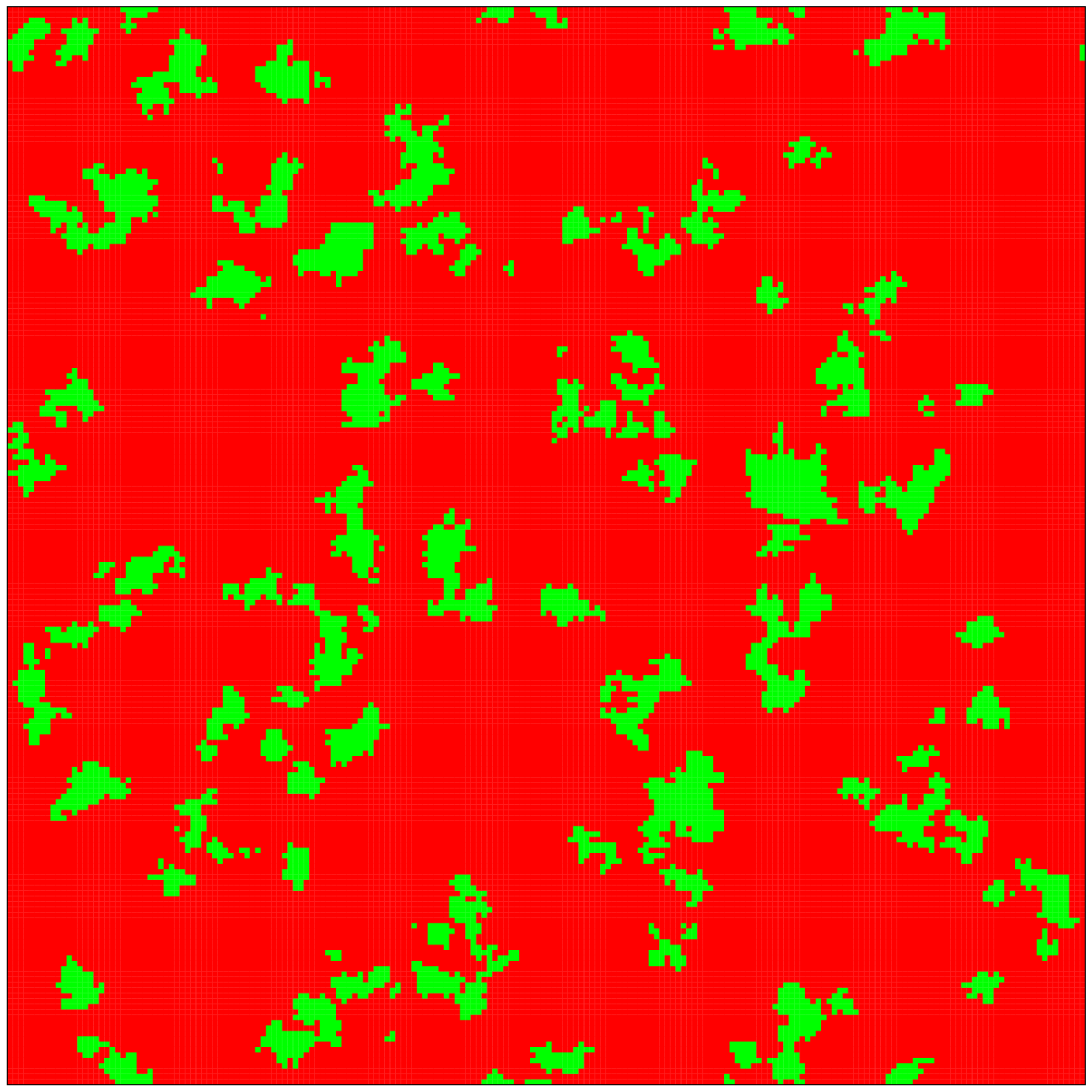}
\includegraphics[width = 4cm]{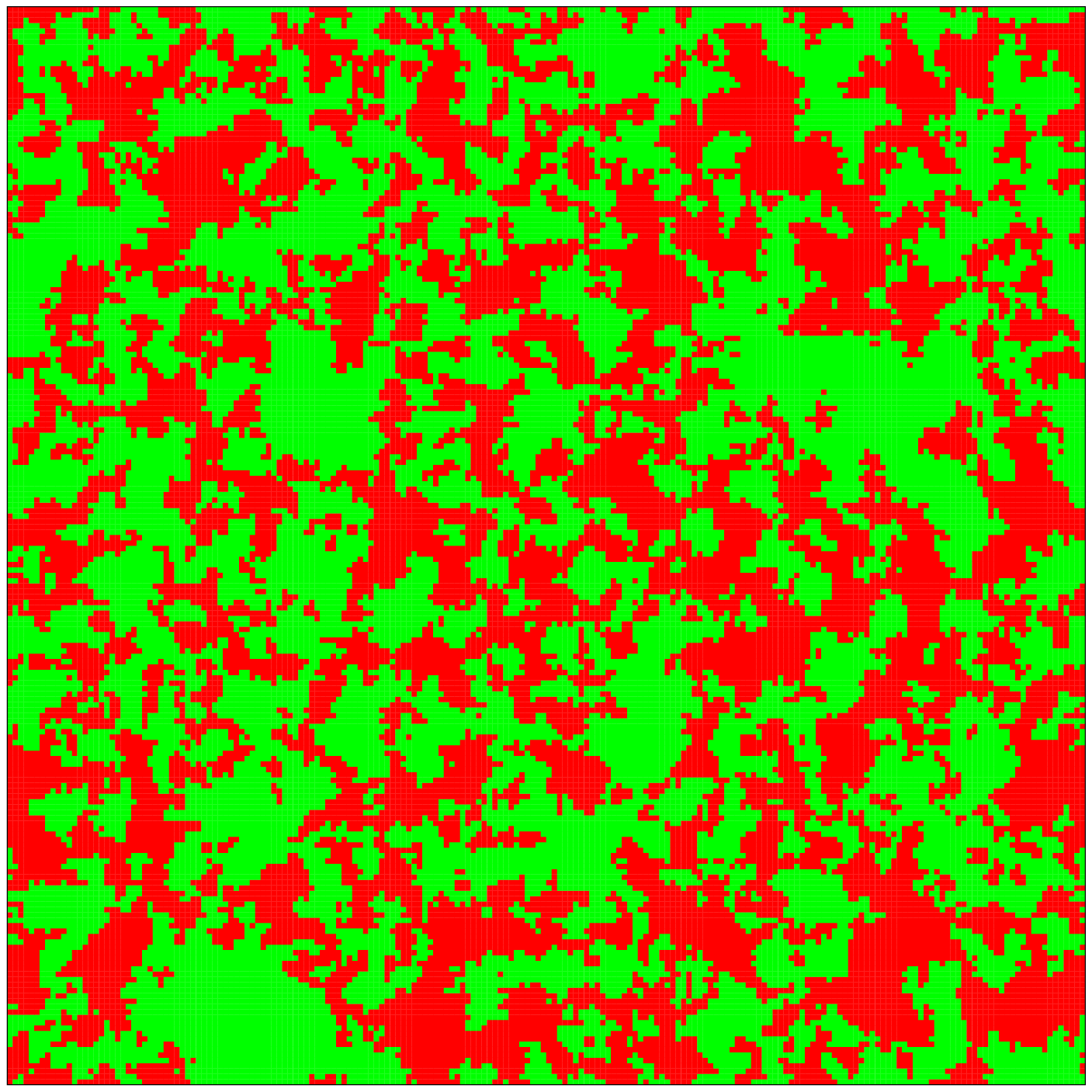}
\caption{\label{snap}Characteristic spatial distributions of the two surviving strategies $i=j=0$ (red) and $i=j+1=1$ (green) for $N=3$ (left) and $N=5$ (right) in the stationary state. While for $N=3$ the less rational players are forced into small isolated clusters, for $N=5$ the distribution of space on the square lattice is practically even.}
\end{center}
\end{figure}

Before presenting the main results, we briefly survey the outcome of the game under well-mixed conditions. In this case, starting from a random distribution, only the strategy cells which fulfill $p<0.5$ and $q>0.5$ will survive. These cells are neutrally stable because none of the offers will be accepted. Accordingly, the system is subject to random drift that results in a fixation to one of the mentioned strategies in a finite population.

Using a spatial system, we first present the outcome of the evolutionary process for $N=3$ and $N=5$. Figure~\ref{time} features the time evolution of the fraction of strategies for $N=5$ (qualitatively similar behavior can be observed for $N=3$). In both cases only two strategies ($i=j=0$ and $i=j+1=1$) survive and coexist. They represent the two most rational options from the available set, both in terms of the offer and the acceptance level. However, while for $N=3$ the $i=j=0$ strategy clearly dominates, for $N=5$ the two strategies share the square lattice almost equally. This is further confirmed by results presented in Fig.~\ref{snap}, where it can be observed that for $N=3$ (left) the $i=j+1=1$ players (green) survive by forming compact isolated clusters, which protects them against the invasion of the more rational players adopting the $i=j=0$ strategy (red). For $N=5$ (right), on the other hand, the characteristic snapshot in the stationary state reflects the equality of power of the two remaining strategies. The trend, being that for larger $N$ the $i=j+1$ strategy becomes superior, is indicative of the results for $N>5$
that we present next.

\begin{figure}
\begin{center}
\includegraphics[width = 8.8cm]{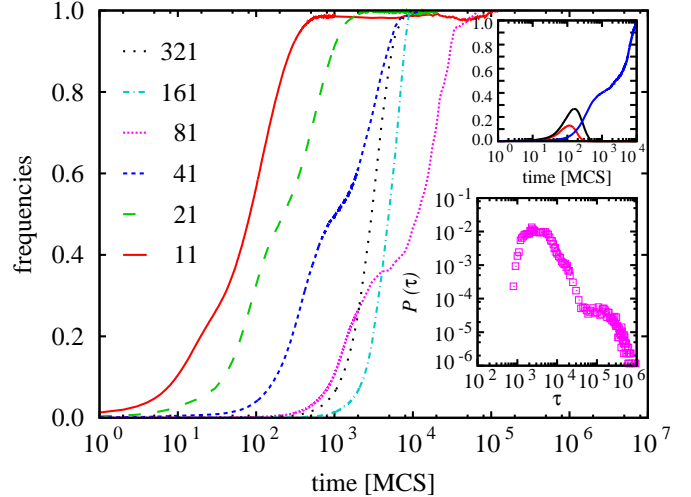}
\caption{\label{more}Time evolution of the fraction of the single surviving strategy for different values of $N>5$ (see legend) and $L=800$. Upper inset shows the time evolution of the three strategies competing the longest, as obtained for $N=41$ and $L=400$. As an illustration of how the fixation time fluctuates, the lower inset shows the probability distribution of the fixation time $\tau$, as obtained for $N=81$ and $L=200$, and averaged over $10^4$ independent realizations of the game.}
\end{center}
\end{figure}

The time evolution of strategies changes dramatically if we increase $N$. Then the system fixates into a homogeneous state where only a single $E_{i,j}$ strategy remains alive. Figure~\ref{more} features the time evolution of the fraction of the only surviving strategy for several different values of $N$, as indicated in the legend. The upper inset reveals how the lastly dominating strategy acquires superiority over the two most persistent opponents. In general, the larger the $N$ the longer the relaxation towards the stationary state, although it is worth noting that the individual fixation times fluctuate heavily even for large $L$. This is illustrated in the lower inset of Fig.~\ref{more}, where the probability distribution of individual fixation times $P(\tau)$ is plotted for $N=81$ and $L=200$. Here the average fixation time is 31290 MCS and the variance is 81140 MCS. Qualitatively similar large fluctuations are characteristic also for other values of $N$ and $L$.

\begin{figure}
\begin{center}
\includegraphics[width = 8cm]{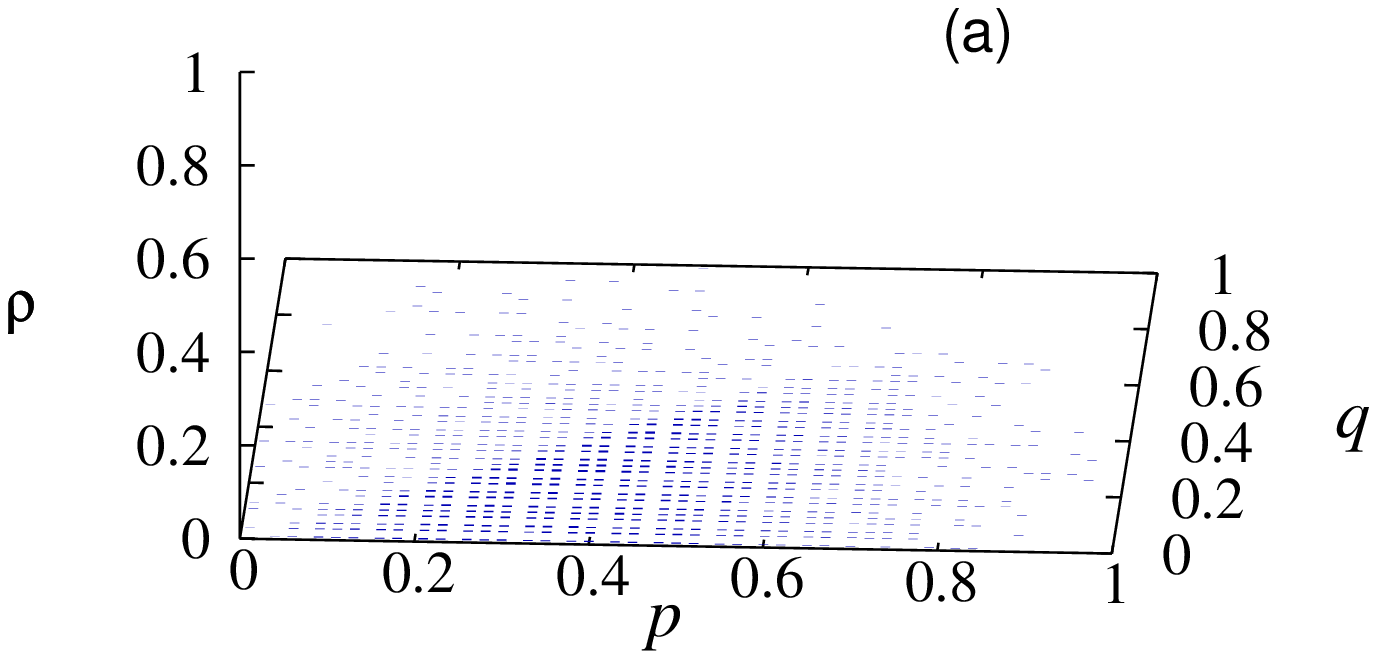}
\includegraphics[width = 8cm]{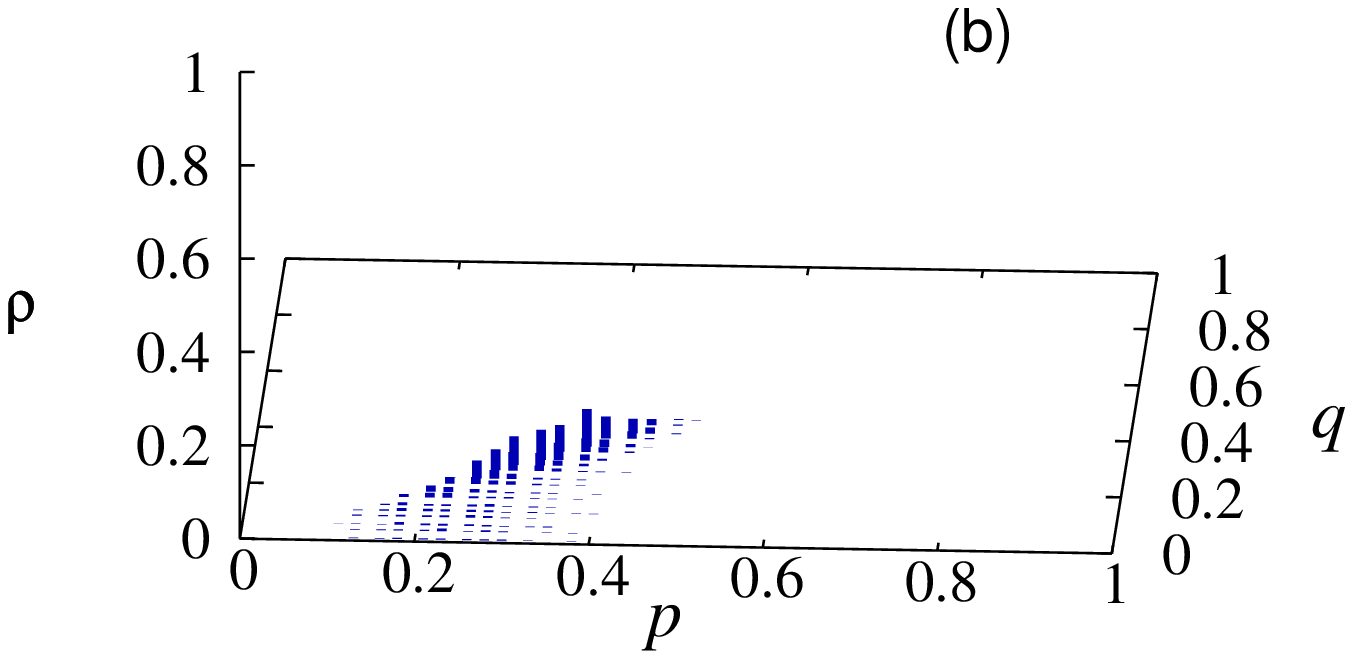}
\includegraphics[width = 8cm]{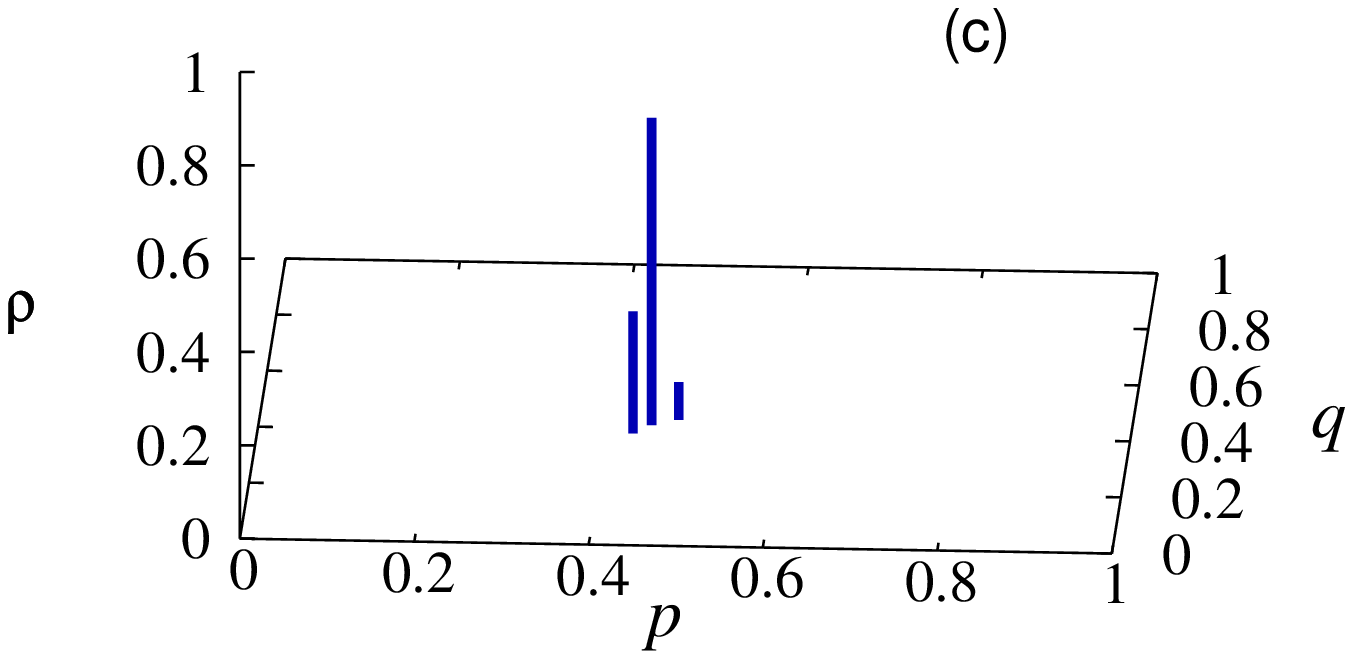}
\includegraphics[width = 8cm]{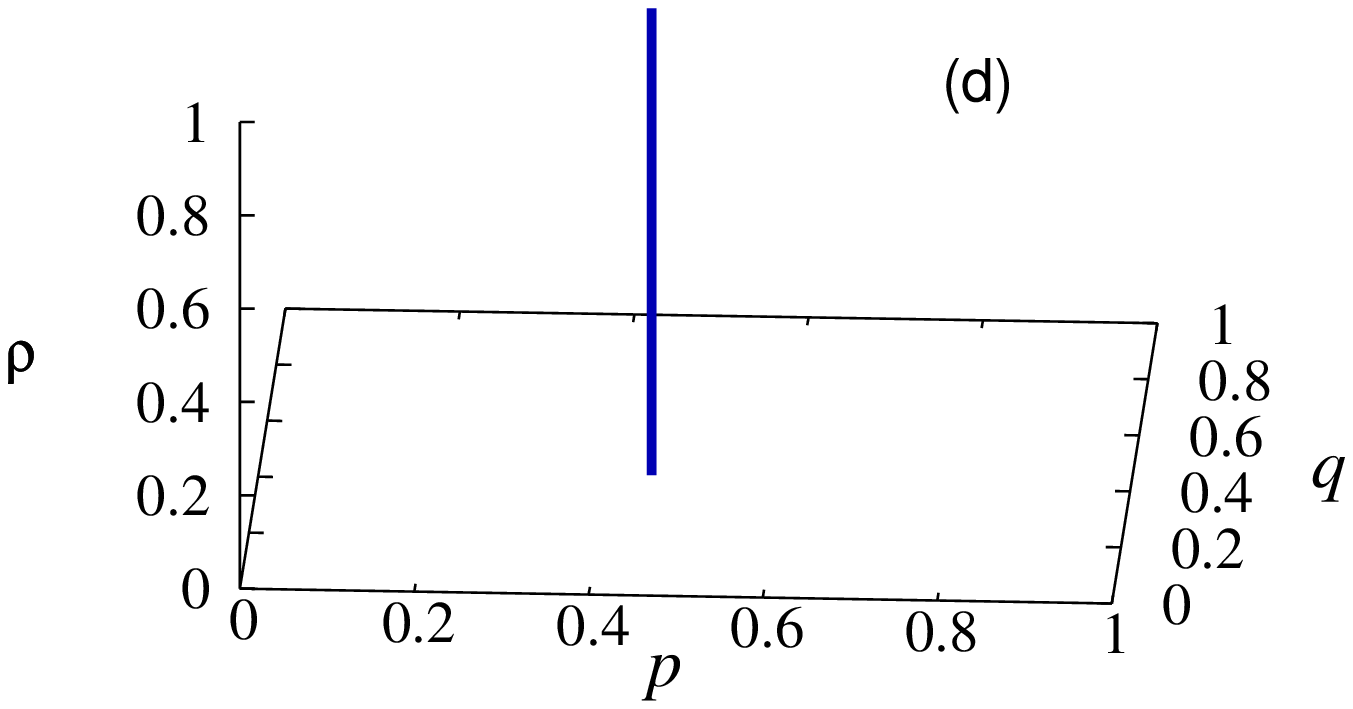}
\caption{\label{dist}Time evolution of the strategy distribution after 5 (a), 50 (b), 5000 (c) and 12000 (d) MCS, as obtained for $N=41$ and $L=400$. First the strategies characterized by $q>p$ become extinct [see panel (a)]. It quickly comes down to determining the victor amongst those strategies that are characterized by $p \approx q < 0.5$ [see panel (b)]. As the evolutionary process matures, only a few of these strategies remain in the system [see panel (c)], upon which lastly a single one survives [see panel (d)].}
\end{center}
\end{figure}

In order to obtain a better understanding of the selection process leading to the dominance of a single strategy, it is instructive to examine the evolution of the strategy distribution over time. Figure~\ref{dist} displays this process for $N=41$ and $L=400$. In the earliest stages of the game, a substantial number of strategies (for example $q>p$, $p>0.5$ and $q>0.5$) die out fast, as illustrated in panels (a) and (b). It is interesting to note that this evolution is in stark contrast to the well-mixed case where strategies from the upper-left region of the $p-q$ plane remain alive. After the extensive decrease in the number of surviving strategies, mainly the empathetic ones ($p \approx q$) survive and continue the competition [see panel (c)]. It is within this part of the evolutionary process that the potential winners (strategies with the higher frequency) exchange consecutively, until finally only one strategy prevails [see panel (d)].

Although at certain stages of the game a strategy may appear as the likely winner, the latter still cannot be foretold conclusively. As indicated by the distribution presented in panel (d), one strategy finally emerges as the victor and occupies the whole lattice ($\rho=1$). Interestingly however, having a closer look at the time evolution of the density of specific strategies at any given $N$ reveals that, during the early stages of the game, several other strategies may occupy a significantly larger portion of the lattice than the strategy that will eventually rise to complete dominance (results not shown). This suggests that the victorious strategy is unable to invade the
whole population at once, but rather that its dominance emerges as a results of an intricate battle between different strategy pairs that cancel each other out beforehand. The battle of initially
successful strategies is illustrated in the upper inset of Fig.~\ref{more}. Revealing such a scenario is possible exclusively due to the consideration of a finite number of competing
strategies, which in turn further supports the important role of local strategy patterns for the outcome of the spatial ultimatum game \cite{szolnoki_ug1}.

The final destination can be different even at the same $N$, yet the difference between $i$ and $j$ is always small, rarely exceeding $2$, and this only for very large $N$. In the majority of cases the victorious strategies differ in dependence on $N$ as follows: $i=j+1=3$ at $N=11$, $i=j+1=8$ at $N=21$, $i=j+1=18$ at $N=41$, $i=j+1=37$ at $N=81$, and $i=j+1=76$ at $N=161$. These results suggest that the evolution of fairness depends sensitively on the multitude of choices players have when posing their ultimatums. In particular, the higher the $N$ the closer the winning strategy will be to the fair one. The evolution of empathy, on the other hand, is much less affected by it.

\begin{figure}
\begin{center}
\includegraphics[width = 8cm]{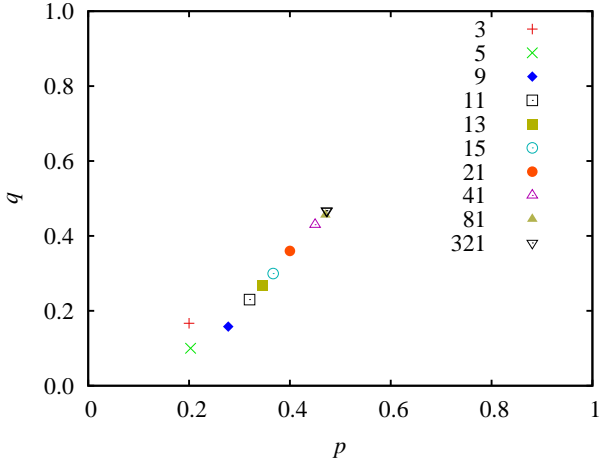}
\caption{\label{fix}The average of victorious $(p,q)$ values, as obtained for different values of $N>5$. For comparison, we also plot the average of surviving strategies for $N=3$ and $N=5$, where stable coexistence can be observed. The trace of destinations shows clearly that the final average $(p,q)$ values converge to the fair $p=q\approx0.5$ solution as we increase $N$.}
\end{center}
\end{figure}

To elaborate on this observation more precisely, we have averaged the victorious $(p,q)$ pairs in dependence on $N$. Results presented in Fig.~\ref{fix} confirm that the larger the $N$ the closer the winning strategy is to the fair $p \approx q =0.5$ outcome. It is worth emphasizing, however, that only in the $N \to \infty$ limit is the fair strategy strictly recoverable. In practice this may be irrelevant, given that the differences become negligible as $N>100$, and can indeed be easily attributed to mutation. Nevertheless, results presented in Fig.~\ref{fix} attest clearly to the fact that, even if the interactions amongst players are restricted, the evolution of fairness requires ample freedom of choice in terms of the selection of $p$ and $q$ values that determine the individual offer and acceptance levels. The evolution of empathy, on the other hand, is much more robust to variations of $N$, given that the fixation points in Fig.~\ref{fix} all fall predominantly in the vicinity of the diagonal.

It is worth mentioning that the above analysis was repeated also for the case when the mutation during the strategy adoption was suppressed by choosing $r_1=r_2=0.5$ for each imitation. We find that for such a setup the system reproduces all the reported main results.

Finally, we would like to stress that the observed fixation to the ``almost fair'' strategies, as well as the tendency that their position on the $p-q$ plain approaches the strictly fair solution as $N$ increases, is not limited to the square lattice interaction topology. In order to confirm this, we have performed simulations on regular small-world networks when $\nu=0.1$ fraction of the links constituting the square lattice was rewired to randomly chosen other players \cite{watts_dj_n98, szabo_jpa04}. This modification of the interaction topology does not affect the main results, indicating that restricted connections, which are representative for all structured populations, represent the most distinctive feature that enables the observations reported in this letter.

In summary, we have studied the spatial ultimatum game with a discrete set of strategies for two types of dynamics based on stochastic imitation of a neighboring strategy. The strategy set included $N^2$ coarse grained strategies covering equidistantly the whole $p-q$ plane. We have shown that for low values of $N$ two strategies can coexist in the stationary state, whereby the survivors are always the two most rational strategies with the lowest possible acceptance level $q$. For larger $N$ only a single strategy survives, which is typically characterized by $i=j+1=(N-1)/2$. Evidently for large $N$ the $p$ and $q$ values of the dominant strategy go to $1/2$, reflecting that the victorious strategy is both fair as well as empathetic. This is significantly different from the mean-field behavior. The quantitative analysis of the evolutionary process indicates that the strategies with $j>i$ ($q>p$) die out quickly, while the strategies with $p \approx q$ remain alive for a long time, yet dependent on $N$. Finally, if $N>5$ one of the $p \approx q$ strategies emerges as the victor, while for $N \leq 5$ we have coexistence as described. Despite of the large size of the system we have used in our simulations, there is some uncertainties in the values of $p$ and $q$ (or $i$ and $j$) for the victorious strategy, which must be related to the small differences in the superiority for the given level of noise.

Despite of the mentioned uncertainties, the presented results indicate clearly that the evolution of fairness is very sensitive to the accuracy in the strategy imitation, which in the present model is characterized by the number of choices the players have for posing their ultimatums. On the other hand, the evolution of empathy seems to be much more robust. For the \textit{Homo emoticus} to shine through, it hence appears we need to have all the freedom possible when engaging into ultimatum bargaining. Such conditions, however, are certainly not always given, which may be one of the reasons the rational \textit{Homo economicus} sometimes gets the best of us. We hope an experimental test will one day be made to shed further light on the importance of discrete strategies in the ultimatum game.

\acknowledgments This research was supported by the Hungarian National Research Fund (grant K-101490) and the Slovenian Research Agency (grant J1-4055).

\end{document}